%% file: main.tex
\documentclass[conference]{IEEEtran}

\input{packages}
\input{shorthands}

\input{cmds}

\def\FIGDIR{./Images}          
\IEEEoverridecommandlockouts

\hypersetup{bookmarks=false}
\hypersetup{draft}

\begin{document}
\makeatletter

\renewcommand\AB@affilsepx{, \protect\Affilfont}
\makeatother

\title{Understanding the Impact of On-chip Communication on DNN Accelerator Performance}
\author[*]{Robert Guirado}
\author[$\dag$]{Hyoukjun Kwon}
\author[*]{Eduard Alarc\'on}
\author[*]{Sergi Abadal\thanks{This work is supported by the grant H2020-863337-WIPLASH.}
}
\author[$\dag$]{Tushar Krishna}
\affil[*]{Universitat Polit\`ecnica de Catalunya}
\affil[$\dag$]{Georgia Institute of Technology}




\maketitle

\begin{abstract}
\input{00-Abstract.tex}
\end{abstract}

\input{01-Introduction.tex}

\input{02-Background.tex}
\input{03-Methodology.tex}

\input{04-Results.tex}

\input{05-Discussion.tex}

\input{06-Conclusion.tex}



\vspace{0.4cm}
\small{
\bibliographystyle{ieeetr}
\bibliography{ref}
}

\end{document}

%% file: packages.tex
\usepackage[usenames,dvipsnames]{xcolor}

\usepackage{setspace}
\usepackage{comment}
\usepackage[most]{tcolorbox}
\usepackage{pifont}

\usepackage{endnotes,xspace,graphicx,fancyvrb,multirow}

\usepackage{paralist}

\usepackage{mathptmx} 

\newcommand{\ignore}[1]{}
\usepackage{fancyhdr}
\usepackage[normalem]{ulem}
\usepackage[hyphens]{url}
\usepackage{microtype}
\usepackage{flushend}
\usepackage[sort,nocompress]{cite}

\usepackage[normalem]{ulem}
\usepackage{float}
\usepackage{authblk}
\usepackage{comment}
\usepackage{caption}
\usepackage{multirow}
\usepackage{subfigure}
\usepackage{color, soul}
\usepackage[nolist,nohyperlinks]{acronym}
\usepackage{upgreek}
\usepackage{graphicx}
\usepackage{dblfloatfix}
\usepackage{epstopdf}
\newcommand{\squishlist}{
 \begin{list}{$\bullet$}
  { \setlength{\itemsep}{0pt}
     \setlength{\parsep}{3pt}
     \setlength{\topsep}{3pt}
     \setlength{\partopsep}{0pt}
     \setlength{\leftmargin}{1.5em}
     \setlength{\labelwidth}{1em}
     \setlength{\labelsep}{0.5em} } }

\newcommand{\squishlisttwo}{
 \begin{list}{$\bullet$}
  { \setlength{\itemsep}{0pt}
     \setlength{\parsep}{0pt}
    \setlength{\topsep}{0pt}
    \setlength{\partopsep}{0pt}
    \setlength{\leftmargin}{2em}
    \setlength{\labelwidth}{1.5em}
    \setlength{\labelsep}{0.5em} } }

\newcommand{\squishend}{
  \end{list}  }

\usepackage{hyperref}

%% file: shorthands.tex
\newcommand{\maestro}{\textsc{MAESTRO}\xspace}

%% file: cmds.tex
\newcommand{\betterparagraph}[1]{{\textbf{#1. }}}

\newcommand{\insertSmallFigure}[2]{
    \begin{figure}[t]
\setlength{\abovecaptionskip}{-1pt}
\setlength{\belowcaptionskip}{-1pt}
        \centering
        \includegraphics[width=0.6\linewidth]{\FIGDIR/#1.pdf}
	\vspace{-1mm}
        \caption{\small #2}
        \label{fig:#1}
    \end{figure}
}

\newcommand{\insertWideFigure}[2]{

    \begin{figure*}[h]
    \setlength{\abovecaptionskip}{-1pt}
    \setlength{\belowcaptionskip}{-4pt}
        \centering
        \includegraphics[width=\textwidth]{\FIGDIR/#1.pdf}
	    \vspace{-6mm}
        \caption{\small #2}
    	\vspace{-4mm}
        \label{fig:#1}
    \end{figure*}
}

%% file: 00-Abstract.tex
Deep Neural Networks have flourished at an unprecedented pace in recent years. They have achieved outstanding accuracy in fields such as computer vision, natural language processing, medicine or economics. Specifically, Convolutional Neural Networks (CNN) are particularly suited to object recognition or identification tasks. This, however, comes at a high computational cost, prompting the use of specialized GPU architectures or even ASICs to achieve high speeds and energy efficiency. ASIC accelerators streamline the execution of certain dataflows amenable to CNN computation that imply the constant movement of large amounts of data, thereby turning on-chip communication into a critical function within the accelerator. This paper studies the communication flows within CNN inference accelerators of edge devices, with the aim to justify current and future decisions in the design of the on-chip networks that interconnect their processing elements. Leveraging this analysis, we then qualitatively discuss the potential impact of introducing the novel paradigm of wireless on-chip network in this context.


%% file: 01-Introduction.tex
\section{Introduction}

\label{sec:intro}

The last decade has witnessed an explosive growth in the development of Neural Network (NN) algorithms both in industry and academia.
Convolutional Neural Networks (CNN) are one of the most successful NNs for popular applications such as image classification~\cite{krizhevsky2012imagenet,vgg},  pose estimation~\cite{vision} or autonomous driving~\cite{tian2018deeptest}, among others.
%
CNNs have been improving their performance over the last years, in part by means of making the NNs larger and deeper, which allowed them to suit more complex forms of data in their design.
The expansion of these networks, often called Deep Neural Networks (DNN), means that they can nowadays reach sizes of millions of parameters \cite{vgg} in some cases,  which inevitably causes a huge computational expense.

As a consequence, the hardware choice to run the inference of those algorithms has evolved to match the new requirements. 
While CPUs provide flexibility and GPUs later offered mass parallelization, they are not specialized hardware for running DNNs and their performance per watt can be improved. 
Moreover, the processing of DNNs is already shifting from the cloud to the edge, hence leading to their wide deployment in devices such as smart home assistants, IoT sensors or autonomous cars. These devices are constrained by strict power envelopes limiting the available hardware resources~\cite{mobilenetv2}.
Therefore, highly tuned architectures specialized for DNNs are required.
Such specialized hardware is generally referred to as DNN accelerators and essentially consists of ASIC architectures optimized for running large NNs.
Several accelerators aiming at ML inference have been released recently \cite{eyeriss, flexflow, maeri}, including novel ideas to boost their efficiency. 


In DNN accelerators, the computational resources are of utmost importance, but the communication is essential as well. 
For instance, in order to efficiently map the NN workloads in the accelerator, data movement is typically leveraged to reuse or parallelize certain computations, which leads to different dataflows or mapping strategies.
Even though data movement is argued to be much less affordable than computation \cite{eyeriss}, most of the designs are focusing on the latter and setting aside the communications, leading to bottlenecks or non-scalable designs. The DNN model dictates the reuse opportunities that can be exploited inside an accelerator. This depends on the data movement, which will be ruled by a dataflow. Moreover, different NN sizes imply unalike data movement inside an accelerator, which requires varying data reuse strategies.

The dataflow space is huge. Therefore, exploring it is a time consuming and critical task towards taking full advantage of the hardware resources with the maximum efficiency.
Consequently, a deep understanding on how these communication approaches affect the computational throughput and runtime of the entire accelerator is needed.
Characterizing the requirements of such NNs becomes, then, essential to understand the impact of the interconnects in the accelerator.

When interconnecting the processing elements of the accelerator among them and with the memory, Networks-on-Chip (NoC) are currently used.
For instance, Eyeriss~\cite{eyeriss} employed hierarchical buses, and MAERI employed a fat-tree and a novel reduction tree~\cite{maeri}.
%
To deliver the data as the dataflow dictates without bottlenecks, NoCs need to be efficient and, as we will see, flexible.
However, NoCs are starting to lag behind and show some limitations, especially when the number of cores increases, as it is happening in general manycore systems~\cite{krishna2013breaking}.
In order to solve this problem, the research community has introduced a novel interconnect paradigm, the Wireless Networks-on-Chip (WNoC), which can address some of the issues traditional NoCs have.
Since DNN accelerators are in essence manycore systems, some advantages can be found if we introduce WNoCs in them.
However, prior work has to be done in order to find the precise dataflows that would benefit from the unique characteristics of WNoCs in DNN accelerators.


In this work, we characterize the NoC bandwidth requirements and the impact of NoC bandwidth on performance in five dataflows and state-of-the-art workloads~\cite{mobilenetv2, resnet}. Precisely, we target inference accelerators on edge devices, where this performance analysis is particularly relevant due to their inherent hardware constraints.
We then explore the design space under these scenarios and reason about the strengths and potential usability of WNoCs in DNN accelerators.



The rest of this paper is organized as follows.
Section II provides background about DNN accelerators, Section III discusses the design space, Section IV presents characterization results, Section V discusses WNoCs, and Section VI concludes.


%% file: 02-Background.tex
\section{Background}
\label{sec:background}

Since CNN is one the most dominant NN class in deployment, most of DNN accelerators focus on CNNs.
Therefore, we also focus on CNNs in this paper. 

\begin{figure}[h]
\vspace{-0.4cm}
\centering
\includegraphics[width=1\columnwidth]{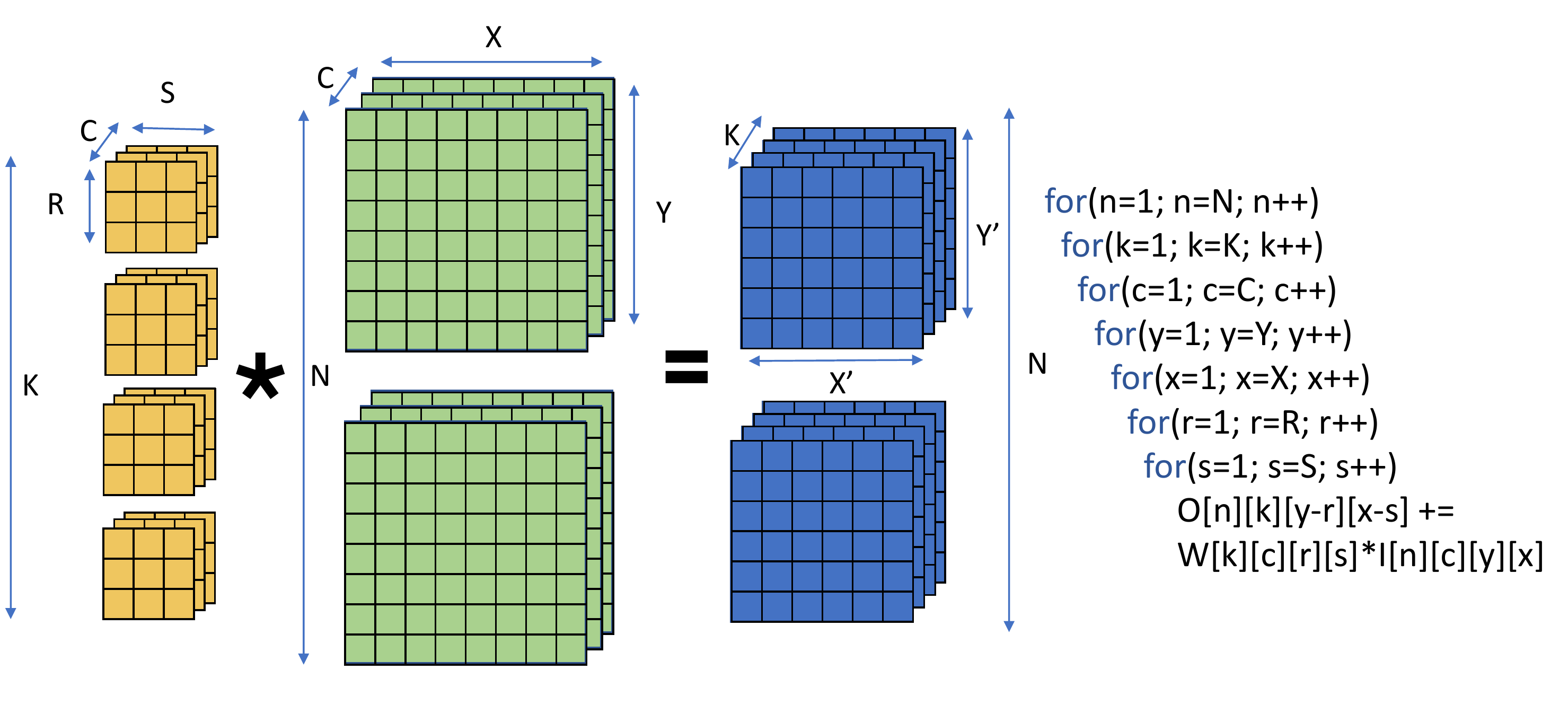}
\caption{Convolution operation}
\label{fig:dataspace}
\vspace{-0.2cm}
\end{figure}

In CNNs, the vast majority of the layers are convolutional (CONV) layers~\cite{diannao}, which essentially perform a discrete convolution operation between the input and the filters or kernels, as~\autoref{fig:dataspace} shows. As a consequence, several designs are using spatial architectures that can take advantage of it to handle efficiently such computations in a tailored manner.


Generally, a DNN accelerator is composed of a memory, an array of Processing Elements (PEs) and a NoC to interconnect the PEs and memory. The PEs fetch inputs and weights from the memory to compute a convolution operation (i.e. multiply and accumulate, MAC), and then send the outputs back to memory. 
The different strategies to sequentially map the NN parameters into the PEs are called dataflows.

When mapping the different dimensions of the NN layers into the PEs, we can spatially or temporally
map each of them, in different orders and sizes, to define our dataflow.
When temporally mapping the data, we iterate in time the different values of such dimension in a PE, while when spatially mapping a dimension, we send the different values to different PEs in our PE array. For instance, dataflows that spatially map the largest dimension of a specific layer will tend to work better in that layer since they facilitate its parallelism. 


\insertSmallFigure{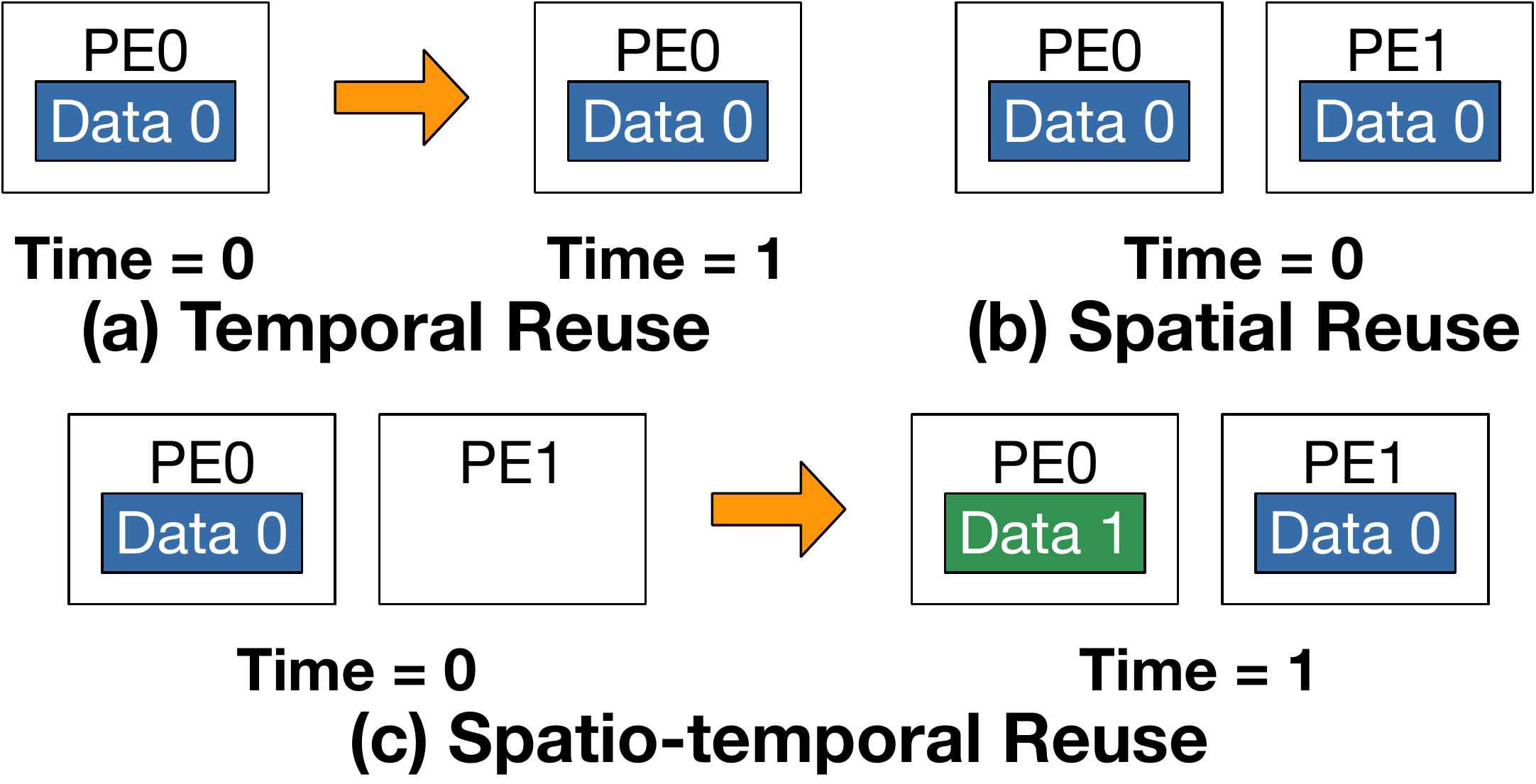}{Three data reuse patterns in accelerators. Temporal reuse is data staging in a buffer within a single PE, spatial reuse is replication over wires (multicast) at the same time, and spatio-temporal reuse is data forwarding from a PE to another PE (e.g., PE0 forwarded data 0 to PE1 in (c)) \vspace{-7mm}}

We can express the dimension space of the convolution operation as a 7D nested for loop, exemplified in~\autoref{fig:dataspace}.
Loop transformations on the 7D loop such as loop interchange and loop tiling lead to completely different dataflow styles, which significantly affects the performance and efficiency of accelerators.
Different dataflows lead to reuse patterns described in~\autoref{fig:DataReuse} and parallelism opportunities whose amount depends on NN layer shapes.
DNN accelerators exploit such opportunities using their available hardware resources: spatial reuse requires multicast ability in the NoC, temporal reuse requires memory hierarchy, and spatio-temporal reuse requires neighbor to neighbor links.




Depending on the temporal data reuse patterns, the following taxonomy was introduced\cite{eyeriss}:
\begin{itemize}
    \item \textbf{Weight Stationary (WS): } WS dataflow refers to a dataflow style that temporally reuses filter weight values in each PE.
    That is, filter weight mapping over each PE changes in the most slowest manner than other tensors.
    NVDLA dataflow style~\cite{nvdla} is a variant of it.
	\item \textbf{Output Stationary (OS): } OS dataflow style tries to accumulate as many partial sums within a PE over time as to reduce the output collection traffic and its cost.
	ShiDiannao~\cite{shidiannao} implements a version of OS dataflow.
	\item \textbf{Row Stationary (RS): } RS dataflow style maps an output row on a set of PEs and accumulates outputs within the row over time.
	RS dataflows spatially reuse filter weight and input activation values and temporally reuse output values across a set of PEs (or, accumulate a full output across a set of PEs). Eyeriss~\cite{eyeriss} implements it.
    \item \textbf{No Local Reuse (NLR): } NLR dataflow does not keep any data stationary, which extremely minimizes buffer sizes but heavily relies on the NoC to move data.
\end{itemize}

Dataflow choices result in different traffic patterns thus requiring different NoC bandwidth, multicast ability, among others.
For example, ShiDiannao~\cite{shidiannao} dataflow requires weight broadcasting and input multicast while NLR dataflow in~\autoref{table:dataflows} unicasts both of weight and input.
However, unicasting dataflow is not always worse than broadcasting dataflow.
The efficiency of dataflows depends on the layer type and shape and the amount of available hardware resources.
That is, each of dataflow styles has strengths and weaknesses, and understanding them is essential to design an efficient accelerator.



%% file: 03-Methodology.tex
\section{methodology}
\label{sec:methodology}

In order to evaluate the performance of the different dataflows and characterize their bandwidth requirements, we use \maestro~\cite{maestro}, an open-source analytical cost model.
\maestro takes three sets of inputs: Dataflow, DNN model, and hardware descriptions written in a specification language.
Analyzing data reuse based on the three sets of inputs, \maestro reports various performance and cost information in layer and network granularity, which includes total latency, buffer access counts, energy consumption, NoC bandwidth requirements, buffer requirements, and others.
\maestro supports not only fundamental DNN operators such as CONV2D and FC, but also modern DNN operators such as depth-wise convolution or transposed convolution.

\betterparagraph{DNN Workloads} We characterize the impact of NoC bandwidth on performance using MobileNetV2~\cite{mobilenetv2} and ResNet50~\cite{resnet} since they provide state-of-the-art efficiency and accuracy for classification applications and include various layer types and shapes.
For layer types, or DNN operators, MobileNetV2 includes CONV2D, depth-wise convolution and point-wise convolution.
Resnet50 includes CONV2D, FC, point-wise convolution, and identity mapping (residual links).
For layer shape, like other classification DNN models, early layers have high resolution (large) activation and shallow filter (small number of input and output channels), and late layers have low resolution (small) activation and deep filter (large number of input and output channels).

\input{Tables/DataflowTable.tex}

\betterparagraph{Characterized Dataflows} ~\autoref{table:dataflows} summarizes the specifications of five dataflow styles we characterize.
Three dataflow styles are based on real accelerators (ShiDiannao~\cite{shidiannao}, Eyeriss~\cite{eyeriss}, and NVDLA~\cite{nvdla} styles) and other two dataflow styles are synthetic dataflows: no-local-reuse (NLR) and weight-stationary (WS).
As shown in~\autoref{table:dataflows}, the five characterized dataflow styles have diverse data reuse strategies, temporal and spatial reuse, loop order, and tile sizes.

\betterparagraph{Hardware Parameters} We target accelerators in edge devices in this work and set up the hardware parameters accordingly.
We model an accelerator with 256 PEs and 256KB SRAM in total for shared global buffer in the accelerator, as well as local buffers in each PE.
We vary the NoC bandwidth from 4B/cycle to 256B/cycle, which translates into 4GB/s to 256GB/s bandwidth range in an accelerator with 1GHz clock.
We enable multicasting of NoC to enable spatial reuse.



%% file: Tables/DataflowTable.tex
\begin{table*}[t]
\centering
\setlength{\abovecaptionskip}{0pt}
\setlength{\belowcaptionskip}{0pt}
\scriptsize
  
\begin{tabular} {|@{~} l @{~}|p{2.5cm}|p{3cm}|p{3cm}|p{2cm}|p{3.0cm}|}
\hline
\multicolumn{1}{|@{~} c @{~}|}{\textbf{Accelerator}} 
& \multicolumn{1}{@{~} c @{~}|}{\textbf{Dataflow Strategy}} 
& \multicolumn{1}{@{~} c @{~}|}{\textbf{Temporal Reuse}} 
& \multicolumn{1}{@{~} c @{~}|}{\textbf{Spatial Reuse}} 
& \multicolumn{1}{@{~} c @{~}|}{\textbf{Loop Order}}  
& \multicolumn{1}{@{~} c @{~}|}{\textbf{Tile Size (K,C,Y,X,R,S)}}  \\
\hline

Example for this work
& No Local Reuse (NLR)
& No data reuse
& No data reuse
& KYXRS\textbf{\color{red}C} 
& (1,1,$|R|$,$|S|$,$|R|$,$|S|$)\\
\hline

Example for this work
& Weight Stationary (WS)
& Weight
& Input (B)
& KCRSY\textbf{\color{red}X} 
& (1,1,$|R|$,$|S|$,$|R|$,$|S|$) \\
\hline

ShiDiannao~\cite{shidiannao} 
& Output Stationary (OS)
& Output
& Input (M) and weight (B)
& KCRS\textbf{\color{red}Y}X/KCRSY\textbf{\color{red} X}
& (1,1,$|R|$,7+$|S|$,$|R|$,$|S|$) \\
\hline

Eyeriss~\cite{eyeriss}
& Row-stationary (RS)
& Weight and output row
& Input (M) and weight (B)
& KC\textbf{\color{red}Y}XRS/KCY\textbf{\color{red}X}R\textbf{\color{red}S}
& (2,2,$|R|$,$|S|$,$|R|$,$|S|$) \\
\hline

NVDLA~\cite{nvdla}
& Weight Stationary (WS)
& Weight
& Input (M) and weight (M)
& \textbf{\color{red}K}CYXRS/\textbf{\color{red}C}KYXRS
& (1,64,$|R|$,$|S|$,$|R|$,$|S|$) \\
\hline

 \end{tabular}
  \caption{\small Dataflow styles we charaterize and their characteristics. In spatial reuse column, "M" and "B" indicate multicast and broadcast, respectively. In loop order column, red texts represent parallelized dimension, and "/" indicates another PE hierarchy level. In tile size column, absolute symbol on a dimension  (e.g., $|X|$) represent the size of the corresponding dimension in a layer. 
  \vspace{-5mm} }
  \label{table:dataflows}
\end{table*}

%% file: 04-Results.tex
\section{Characterization Results}
\label{sec:results}

\insertWideFigure{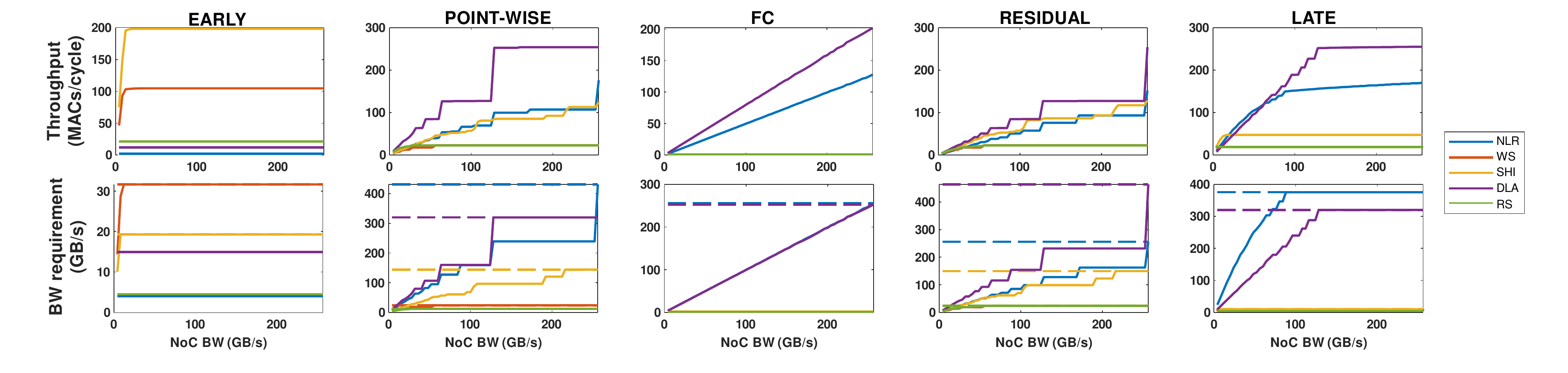}{Resnet50 analysis. Dotted lines represent the peak bandwidth requirements for each dataflow style.}
\insertWideFigure{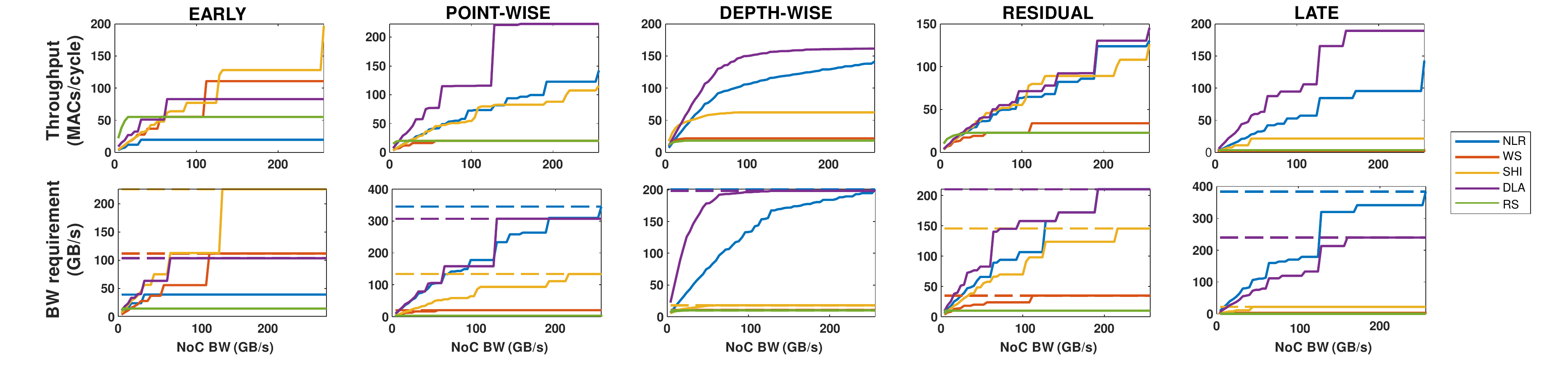}{MobileNetV2 analysis. Dotted lines represent the peak bandwidth requirements for each dataflow style.}

We present the impact of NoC bandwidth on throughput and bandwidth requirement over Resnet50 and MobileNetV2 in~\autoref{fig:ResNet} and~\autoref{fig:MobileNet}.
We classify each layer of the DNN models into five classes: early layer, point-wise convolution, fully-connected layer, residual links, and late layers.
We observe that each dataflow provides different roofline thoughputs and requires different NoC bandwidth.

\betterparagraph{Roofline Throughput} 
Roofline throughput depends on the maximum degree of parallelization in a given set of target layer dimension, dataflow, and PE array size.
First, when the dataflow parallelizes over a layer dimension smaller than number of PEs, PEs can be underutilized.
For example, NLR dataflow style parallelizes over input channel dimension. 
When we run an early layer with only three channels (e.g., CONV1 in Resnet50), only three PEs can be utilized so the maximum throughput is three MACs per cycle.

Also, when the number of PEs does not cover the entire parallelized dimension, the throughput is restricted by the total number of PEs (computation bounded). 
For example, late layers in CONV5 of Resnet50 have more than 512 channels.
Since the evaluated edge accelerator has only 256 PEs, the maximum throughput is bounded to 256 MACs per cycle, as shown in the late layer column in~\autoref{fig:ResNet} and~\autoref{fig:MobileNet}.

NVDLA style dataflow provides the highest roofline throughput in all the layer types other than early layers. This is because the parallelization is over input and output channels, but early layers have a small number of channels compared to the activation size.
However, ShiDiannao and row stationary dataflow styles provide overall low throughput except early layers because they parallelize computation over activations.

For fully connected layers, since the activation size is analogous to the filter size, and the layer shape is extremely narrow and deep, dataflow styles that parallelize over activation can utilize limited number of PEs (e.g., only 1 PE is utilized in row stationary style dataflow).
That is, for such a dataflow, the accelerator designer or a programmer needs to provide alternative processing style for fully connected layers.
Both fully connected layers and residual links require larger amount of bandwidth compared to other layer types because their limited amount of data reuse implied by the operation.

\betterparagraph{Peak and Average Bandwidth Requirements}
Most of the accelerators adopt double buffering techniques to hide the communication latency, which lowers the processing delay of a computation tile from $Delay_{compute} + Delay_{communication}$ to Max($Delay_{compute}, Delay_{communication}$).
That is, data distribution of the next computation tile is performed while a PE array processes its previous tile.
The average bandwidth requirements we plot in~\autoref{fig:ResNet} and~\autoref{fig:MobileNet} reflect such aspects.

Note that the communication delay is for the entire PE array; if 32 data points need to be distributed for a new computation tile while the NoC bandwidth is 12 bytes per cycle, the communication delay is $ceil(32/12) =3$ cycles.
Therefore, increased NoC bandwidth decreases communication delay in a discrete manner.
This leads to discrete increments of throughput over bandwidth when the throughput is communication-bound (or, when the NoC bandwidth is less than peak bandwidth).
We can observe such behaviors in the second row in~\autoref{fig:ResNet} and~\autoref{fig:MobileNet}.

The interval of discrete throughput increment implies the performance sensitivity towards NoC bandwidth.
We observe that the sensitivity depends on both layer types and dataflow styles.
For example, NVDLA dataflow style has large NoC banwidth intervals for throughput increases compared to other dataflows in point-wise convolution operations, as shown in the second column of~\autoref{fig:ResNet} and~\autoref{fig:MobileNet}.
However, NVDLA has relatively lower sensitivity in early layers.

The average bandwidth requirement line summarizes such aspects and provides useful insights:
(1) Depending on the dataflow and layer type, providing more NoC bandwidth can be futile (flat regions in average bandwidth lines)
(2) For some combinations of dataflow and layer type (e.g., NLR in FC layer in~\autoref{fig:ResNet}), NoC bandwidth is critical for performance.

%% file: 05-Discussion.tex
\section{Discussion}
\label{sec:discussion}

While traditional NoCs have been introduced to easily interconnect different cores inside a package, DNN accelerators are starting to embody more and more number of processing elements.
This has led to a situation in which they face some challenges such as scalability problems, lack of flexibility for the mappings, lack of efficient broadcast support, limited bandwidth and area and power over-consumption.
As our analysis shows, dataflow flexibility and bandwidth alone have a huge impact in the throughput of an accelerator, and therefore having an interconnect -such as WNoC- that provides solutions for these challenges is a promising approach to boost the performance of DNN accelerators.

WNoC allow to virtually map different topologies on-demand at every cycle, allowing the required adaptability on the dataflows and scaling the designs to thousands of PEs.
Studies in this topic \cite{onthearea,replica} have shown that WNoCs have great potential and may fulfill these requirements.

%% file: 06-Conclusion.tex
\section{Conclusion}
\label{sec:conclusions}

In this work, we characterized the impact of NoC bandwidth on throughput and bandwidth requirement of DNN dataflows over state-of-the-art DNN models with diverse layer types and shapes.
From the characterization results, we can observe that both flexibility and sufficient bandwidth are indispensable requirements for the NoC inside accelerators, which is challenging for traditional NoCs to provide.
However, emerging interconnection technologies such as WNoC may be able to deliver such desired features, therefore turning them into promising candidates for the implementation of on-chip networks within DNN accelerators.



\vspace{0.1cm}